\documentclass[preprint]{revtex4}
\usepackage{footnote}
\usepackage[dvips, bookmarks, colorlinks=true, plainpages = false,
citecolor = green, urlcolor = blue, filecolor =
blue,hypertex,hyperref]{hyperref}
\makeindex
\usepackage{amsmath}
\usepackage{amsxtra}
\usepackage{amstext}
\usepackage{amssymb}
\long\def\symbolfootnote[#1]#2{\begingroup\def\thefootnote{\fnsymbol{footnote}}
\footnote[#1]{#2}\endgroup}

\newcommand{\spa}{\ , \ \ }

\begin{document}
\title{Absorption Cross Section of Smeared D$3$-Brane On A Circle}
\author{Anastasios Psinas}
\affiliation{Department of Physics,\\ University of Patras \\
GR-26500, Patras, GREECE} 
\begin{abstract}




The evaluation of the absorption cross section of a massless scalar
field propagating on a non-extremal black D$3$-brane smeared on a
circle is considered. The solution to the scalar field equation of
motion at high temperature is obtained in terms of the parameters
$\lambda=3\omega/4\pi T$ and $k$, which denote the frequency and the
momentum along the circle respectively while $T$ denotes the
temperature of the black brane. Based on a perturbative scheme,
higher order temperature corrections to the scalar absorption cross
section are computed. Further, interesting analogies with the cross
section of known black branes solutions are discussed.

\end{abstract}
\maketitle

\section{Introduction}

According to Bekenstein-Hawking
\cite{Bekenstein:1973ur},\cite{Bekenstein:1974ax},\cite{Hawking:1974sw}
seminal work the gravitational entropy of a Schwarzschild black hole
in proper units, is proportional to the area spanned by its horizon
namely, $S=A/4$. Although classically such black holes are stable
this is not the case when quantum mechanical effects are taken into
account. One can picture Hawking radiation as particles which are
absorbed by the black hole while at the same time radiation is
emmited to its environment. A very interesting result
\cite{Das:1996we} states that the absorption cross section
probability of a massless scalar field embedded on a spherically
symmetric black hole in $p+2$ dimensions is simply given by the area
of the horizon. Similar studies have been performed for more general
higher dimensional spacetimes \cite{Das:1996wn,Gubser:1996xe}
leading to interesting results regarding the low frequency behavior
of the absorption probability of neutral scalars. Additionally, it
is known \cite{Klebanov:1997kc}, that for non-dilatonic extremal
$D$-branes in the s-wave approximation $\mathcal{P}_{abs}=0$ as
$\omega$ approaches zero. An extensive treatment of Hawking emission
rates even for higher partial waves in effective string models can
be found in
\cite{Maldacena:1996ix},\cite{Gubser:1996zp},\cite{Callan:1996tv},\cite{Klebanov:1997gt},
\cite{Dowker:1997vj},\cite{Hawking:1997nw},
\cite{Krasnitz:1997gn},\cite{Mathur:1997et},\cite{Gubser:1997qr}.
The importance of such calculations including non-extremal black
D-branes extends also to evaluating transport properties of gauge
field theories at strong coupling \cite{Policastro:2001yc}.

Most calculations that have been done so far including black
D-branes have a horizon of spherical topology i.e. $S^{P}$. In this
paper we will focus on a different type of branes the so called
smeared D$p$-branes. Smearing a brane along one of its transverse
directions creates an electrically charged brane provided it has a
direction with a translational symmetry along which the original
brane is neutral. As is in ordinary D-branes one can define a near
horizon limit making a connection to a dual supersymmetric
Yang-Mills gauge theory living in a lower dimensional spacetime.
Interesting phenomena have emerged out of studying such branes such
as a Gregory-Laflamme \cite{Gregory:1993vy,Gregory:1994bj} type
instability that has been observed and extensively investigated. For
more details on the classical and thermodynamic stability of smeared
branes consult
\cite{Gubser:2000ec},\cite{Harmark:2002tr},\cite{Ross:2005vh},\cite{Harmark:2005jk}
and references therein.

In what follows we shall give the details of the evaluation of the
absorption cross section of a scalar field on a D$3$-smeared brane.
This calculation is particularly involved since the metric is such
that introduces a non zero momentum along the smeared direction.
Given that, the solution to the scalar wave equation is obtained as
a double perturbative expansion on $\omega/T$ and $k$ in the high
temperature regime\footnotemark[1]\footnotetext[1]{For an extensive
analysis on the low temperature regime for near extremal branes see
\cite{Policastro:2001yb}}. According to our analysis we obtain the
general expression of $\sigma_{abs}$ for higher partial waves to the
zeroth order in $\lambda$ and $k$ (as dictated by the perturbative
scheme applied on the scalar field), while higher order corrections
in $\lambda$ are carried out in the s-wave approximation.

\section{Smeared black D$p$-branes} We begin by presenting some
basic aspects of smeared d-branes following closely
\cite{Harmark:2005jk}. The way to obtain a smeared D-brane out of
black string solution involves a series of steps. One starts with
the following 11-dim uplift of a black string metric initially
defined in $10-p$ dimensions

\begin{equation}
ds^2_{11}=-fdt^2+dz^2+f^{-1}dr^2+r^2d\Omega^2_{7-p}+\sum^{p}_{i=1}dx^2_{i}+dy^2,\\f=1-\frac{r_0^{6-p}}{r^{6-p}}\label{1}
\end{equation}
where $x_i$ stands for $p$ flat directions, and $y$ denotes the
direction that we will apply a IIA reduction (S- duality
transformation). Next, one applies a boost along the y-direction
with a boost parameter $\alpha>0$ so that charge is assigned to our
solution. The final step is to perform a T-duality on every $x_i$
direction so that one finally obtains a non-extremal D$p$-brane
along the z-direction

\begin{equation}
\label{2}
\begin{array}{c}
ds^2 = H^{-1/2} \left( - f dt^2 + \sum_{i=1}^p dx_i^2 \right) +
H^{1/2} \left( f^{-1} dr^2 + dz^2  + r^2 d\Omega_{7-p}^2 \right) \ ,
\\[5mm]
H = 1 +  \frac{R^{6-p}}{r^{6-p}} \spa e^{2\phi} = H^{\frac{3-p}{2}}
\spa  A_{01\ldots p} =  \coth \alpha \, ( H^{-1} - 1 ) \spa
R=r_0{\sinh{\alpha}}^{\frac{2}{6-p}}  \,
\end{array}
\end{equation}
where the metric is expressed in the string frame. One can also
define the near-extremal limit of the above metric where details can
be found in \cite{Harmark:2005jk}. Close inspection of the line
element Eq.~(\ref{1}) reveals that it has a non-trivial spherical
topology $S^1\times{S^{7-p}}$. This in particular has an impact on
the evaluation of the scalar wave function $\Phi$ by assigning a
wave number (see $k$) in $\Phi$ even in the s-wave approximation. We
shall expand on that by giving more details in the following
section.

The thermodynamics of the spacetime at hand is shown to exhibit
interesting properties such as a Gregory-Laflamme type instability.
In the case examined though our main concern is to study scalar
instead of gravity perturbations. Starting with the entropy (which
is proportional to the area's horizon), is easy to prove that


\begin{equation}
S=L\frac{\Omega_{7-p}}{16\pi G}V_{p}r_0^{7-p}\cosh\alpha\label{3}
\end{equation}
where $V_{p}$ is the world-volume of the brane, and with $L$ we
denote the circumference of the circle of the smeared direction. We
note, that in the latter part of the paper we show, that the
absorption probability of the scalar is proportional but not equal
to the area of the horizon of the black brane in the low frequency
regime. The temperature ascribed to the black brane is obtained
through implementing
$T=\frac{1}{4\pi}|dg_{tt}/dr|\sqrt{-g^{tt}g^{rr}}$ evaluated at the
location of the horizon $r=r_0$
\begin{equation}
T=\frac{6-p}{4\pi r_{0}\cosh\alpha}\label{4}
\end{equation}
Before advancing further, it would be useful to point out that the
evaluation of the absorption probability will be performed for large
values of the boost factor which basically translates to working in
the near extremal limit i.e. $R>>r_0$.

\section{Scalar Wave Equation}

The main task in this section is to solve the wave equation of a
massless scalar field $\Phi$ which its dynamics is driven by the
smeared D$3$-brane, in other words one has to solve
\begin{equation}
\Box{\Phi}=0\label{5}
\end{equation}
Given the isometries of the brane, it is natural to consider an
ansatz that reads
\begin{equation}
\Phi=e^{i(\omega t+kz)}\phi(r)Y\label{6}
\end{equation}
where $\omega$ and $k$ stand for the frequency and the momentum
along the smeared dimension respectively and $Y$ denotes the higher
dimensional spherical harmonics on the $(7-p)$ sphere. In
$p$-dimensions and working in the Euclidean frame the wave operator
acts in the wave function as follows (s-wave approximation)
\begin{equation}
\Box_{E}\Phi=(\omega^2H^{\frac{7-p}{2}}f^{-1}-k^2H^{-\frac{p+1}{8}})\Phi+r^{p-7}H^{-\frac{p+1}{8}}\partial_{r}({r^{7-p}f\partial_{r}{\Phi}})\label{7}
\end{equation}
A similar although not identical equation can be written in the
string frame. However, in our analysis we are restricted on smeared
D$3$-branes so both frames correspond to the same wave equation. In
addition, if higher partial waves are taken into account the wave
equation for $p=3$ reduces down to
\begin{equation}
\Big(\omega^2H^{\frac{1}{2}}f^{-1}-k^2H^{-\frac{1}{2}}-l(l+3)r^{-2}H^{-\frac{1}{2}}\Big)\Phi+r^{-4}H^{-\frac{1}{2}}\partial_{r}({r^{4}f\partial_{r}{\Phi}})=0\label{8}
\end{equation}
Solving Eq.~(\ref{8}) requires separating the entire region of
spacetime in a far and a near horizon region in which the obtained
solutions must be 'stretched' and matched in an intermediate domain.
The above procedure is a very standard one, usually implemented when
it is impossible to obtain a full solution in the domain where
$\Phi$ is defined.
\subsection{Far Region Analysis}
Before proceeding further in our analysis and for later convenience
is better if we adopt $\rho=\omega r$ and $\rho_0=\omega r_0$. Then
Eq.~(\ref{8}) reads

\begin{eqnarray}
\frac{\partial^{2}\phi}{\partial\rho^{2}}+
\Big(4+\frac{3\rho_0^{3}}{(\rho^{3}-\rho_0^{3})}\Big)\frac{1}{\rho}\frac{\partial{\phi}}{\partial{\rho}}
+\Big(\frac{\rho^{3}(\rho^{3}+(\omega
R)^{3})}{(\rho^{3}-\rho_0^{3})^2}-\frac{l(l+3)\rho}{\rho^{3}-\rho_0^{3}}-\frac{(k^2/\omega^2)\rho^{3}}{\rho^{3}-\rho_0^{3}}\Big)\phi=0\label{9}
\end{eqnarray}
In the outer region, defined as $\rho>>\rho_0$ and $\rho>>(\omega
R)^2$, one can determine the asymptotic from of the radial part of
the wave function in terms of the rescaled radial distance $\rho$.
Thus, the wave equation turns out to be

\begin{equation}
\frac{\partial^2\phi}{\partial\rho^2}+\frac{4}{\rho}\frac{\partial\phi}{\partial\rho}+\Big(\frac{\omega^2-k^2}{\omega^2}-\frac{l(l+3)}{\rho^2}\Big)\phi=0\label{10}
\end{equation}
 Eq.~(\ref{10}) is easily solved in terms of the Bessel functions of
 the first and the second kind

\begin{equation}
\phi(\rho)=\alpha\rho^{(\frac{-3}{2})}J_{l+\frac{3}{2}}\Big(\frac{\sqrt{\omega^2-k^2}}{\omega}\rho\Big)+\beta\rho^{(\frac{-3}{2})}Y_{l+\frac{3}{2}}\Big(\frac{\sqrt{\omega^2-k^2}}{{\omega}}\rho\Big)\label{11}
\end{equation}
where $\alpha, \beta$ are constants whose value is determined by the
asymptotic behavior of the scalar and by proper matching in the
intermediate region. Let us be more specific by recalling that the
above solution described by Eq.~(\ref{11}) is valid for big values
of $\rho$. We can stretch somewhat the solution by trying to extend
it in the so called matching region, $\rho_0<<(\omega
R)^2<<\rho<<1$, which yields
\begin{equation}
\phi(\rho)\simeq{\frac{\alpha}{\Gamma(l+\frac{5}{2})}\Big(\frac{\sqrt{\omega^2-k^2}}{2\omega}\Big)^{l+\frac{3}{2}}\rho^{l}-\beta\frac{\Gamma(l+\frac{3}{2})}{\pi}\Big(\frac{2\omega}{\sqrt{\omega^2-k^2}}\Big)^{l+\frac{3}{2}}\rho^{-l-3}}\label{12}
\end{equation}
However, given that $\omega R<<1$ we must impose $\beta=0$. Finally,
for consistency reasons we demand the condition $\omega^2>k^2$ to
hold.
\subsection{Inner Region Solution}
The wave function in the inner region, $\rho_0<<\rho<<1$, is easy to
obtain by substituting $x=\rho_0/\rho$, in which case Eq.~(\ref{9})
transforms to
\begin{equation}
\frac{\partial^2\phi}{\partial
x^2}-\frac{2+x^3}{x(1-x^3)}\frac{\partial\phi}{\partial
x}+\Big(\frac{\frac{\rho_0^2}{x^3}+\frac{\omega^3R^3}{\rho_0}}{x(1-x^3)^2}-\frac{l(l+3)}{x^2(1-x^3)}-\frac{k^2\rho^2/\omega^2}{x^4(1-x^3)}\Big)\phi=0\label{13}
\end{equation}
In the inner region the $\rho^2/x^3$ term can be neglected when
compared to $\omega^3R^3/\rho_0$, resulting in
\begin{equation}
\frac{\partial^2\phi}{\partial
x^2}-\frac{2+x^3}{x(1-x^3)}\frac{\partial\phi}{\partial
x}+\Big(\frac{\lambda}{x(1-x^3)^2}-\frac{l(l+3)}{x^2(1-x^3)}-\frac{k^2\rho^2/\omega^2}{x^4(1-x^3)}\Big)\phi=0\label{14}
\end{equation}
where $\lambda\equiv{3\omega/4\pi T}$. The definition of $\lambda$
is established through a direct implementation of Eq.~(\ref{4}) in
the near extremal regime.

It is easy to verify that Eq.~(\ref{14}) cannot be solved
analytically. This fact necessitates the need to apply some kind of
approximation scheme allowing us to study the behavior of the wave
function as we vary $x$. Hence, the presence of the $\lambda$
parameter is crucial into distinguishing whether one seeks a
low-temperature ($\lambda>>1$) \cite{Policastro:2001yb} or a
high-temperature ($\lambda<<1$) treatment of the problem. In our
case we shall focus on the latter approximation scheme.

In the region close to the horizon ($x\simeq{1}$), is easy to show
that $\phi(x)\sim{(1-x)}^{\pm{\frac{i\lambda}{3}}}$. However, only
the negative solution is kept since it corresponds to the incoming
wave at the horizon. An improved ansatz is needed when moving away
from $x=1$ in which case an additional interpolating function is
included
\begin{equation}
\phi(x)=(1-x)^{\sigma}F(x)\label{15}
\end{equation}
where $\sigma=-i\lambda/3$. By plain substitution of Eq.~(\ref{15})
into Eq.~(\ref{14}) and after a long and tedious calculation one
gets
\begin{multline}
\frac{\partial^2F}{\partial
x^2}+\Big(\frac{2\sigma}{1-x}+\frac{(2+x^3)}{x(1-x^3)}\Big)\frac{\partial
F}{\partial x} \\
+\Big(\frac{\sigma(\sigma-1)}{(1-x)^2}+\frac{(2+x^3)\sigma}{x(1-x)(1-x^3)}+\frac{\lambda^2}{x(1-x^3)^2}-\frac{l(l+3)}{x^2(1-x^3)}-\frac{k^2r_0^2}{x^4(1-x^3)}\Big)F=0\label{16}
\end{multline}

As is easily observed, Eq.~(\ref{16}) is very cumbersome possessing
no solution in closed form. Moreover, one sees that there are two
parameters involved. Based on that, we wish to apply a double
perturbative expansion on $\lambda,k$ which are taken to be small.
Hence, $F(x)$ is decomposed as
\begin{equation}
F(x)=F^{(0)}+\lambda F^{(1)}(x)+\lambda^2
F^{(2)}(x)+k^2G^{(2)}(x)+...\label{17}
\end{equation}
Studying the behavior of $F^{(0)}$ (zeroth order to $\lambda$ and
$\tilde k$) around $x=0$ one gets two solutions ($x^{-l},x^{l+3}$)
which lead to the ansatz
\begin{equation}
F^{(0)}(x)=x^{(l+3)}F_1(x)+x^{-l}F_2(x)\label{18}
\end{equation}
The unknown functions $F_1, F_2$, are determined by substituting
Eq.~(\ref{18}) into Eq.~(\ref{16}) while neglecting terms of higher
order in $\lambda,\tilde k$. For completeness we provide the
differential equation that $F_1$ obeys
\begin{equation}
y(1-y)\frac{d^2F_1}{d
y^2}+\Big((2+\frac{2l}{3})-(3+\frac{2l}{3})y\Big)\frac{dF_1}{dy}-(1+\frac{l}{3})^2F_1=0\label{19}
\end{equation}
where the following change of variables $y=x^3$ is made. It is easy
to cheque \cite{Tables of Integrals}, that Eq.~(\ref{19}) is a
hypergeometric equation of the form
$F_1(x)=_{2}\mathcal{F}_{1}(1+\frac{l}{3},1+\frac{l}{3};2+\frac{2l}{3};x^3)$.
Following a similar analysis we conclude that
$F_2(x)=_{2}\mathcal{F}_{1}(-\frac{l}{3},-\frac{l}{3};-\frac{2l}{3};x^3)$.
\section{Absorption Cross Section Evaluation}
We are almost ready to advance towards the main goal of the paper
that is evaluating the absorption cross section probability of the
scalar field. There are two distinct cases one has to examine
separately. We begin with the first case where $l\neq{3n}$,
$n=1,2,3,..., $. We would like to have a regular solution at the
horizon, $x=1$, that is to say free of any logarithmic
singularities. It turns out that
$F^{(0)}(x)=x^{l+3}F_1(x)+x^{-l}D_{l}F_2(x)$, where the constant
$D_{l}$ is defined as
\begin{equation}
D_{l}=-\frac{\Gamma(2+\frac{2l}{3})\Gamma^2(-\frac{l}{3})}{\Gamma(-\frac{2l}{3})\Gamma^2(1+\frac{l}{3})}\label{20}
\end{equation}
Upon 'stretching' the solution described by Eq.~(\ref{18}) to the
matching region we recover
\begin{equation}
\phi(\rho)\simeq{D_{l}\rho_{0}^{-l}\rho^{l}}\label{21}
\end{equation}
Hence, Eq.~(\ref{12},\ref{21}) when compared to the intermediate
region ($\beta=0$) give
\begin{equation}
\alpha=2^{(l+\frac{3}{2})}\Gamma(l+\frac{5}{2})D_{l}\Bigg(\frac{\sqrt{\omega^2-k^2}}{\omega}\Bigg)^{(-l-\frac{3}{2})}\rho_0^{-l}\label{22}
\end{equation}
Additionally, the incoming wave from infinity reads
\begin{equation}
\phi_{\infty}(\rho)=\sqrt{\frac{2}{\pi}}\frac{\alpha}{2}\Bigg(\frac{\omega^2}{\omega^2-k^2}\Bigg)^{\frac{1}{4}}\rho^{-2}e^{-i(\frac{\sqrt{\omega^2-k^2}}{\omega}\rho-\frac{(l+2)\pi}{2})}\label{23}
\end{equation}
The wave function very close to the vicinity of the horizon reads
\begin{equation}
\phi(\rho)=\Big(1-\frac{\rho_0}{\rho}\Big)^{-\frac{i\lambda}{3}}\label{24}
\end{equation}
In general, the absorption probability is computed through
\begin{equation}
P(l)\equiv{\frac{(\sqrt{-g}g^{rr})_{h}\Big(\phi^{*}_{h}\frac{\partial
\phi_{h}}{\partial
r}-c.c\Big)|_{r=r_0}}{(\sqrt{-g}g^{rr})_{\infty}\Big(\phi^{*}_{\infty}\frac{\partial
\phi_{\infty}}{\partial r}-c.c\Big)|_{r=r_0}}}\label{25}
\end{equation}
where all wave functions at the horizon and at infinity (as the
subscripts indicate) are evaluated at the location of the horizon
$r=r_0$. After a careful evaluation of the flux fraction we conclude
that the absorption probability is simply given by
\begin{equation}
P(l)=\frac{2\pi\lambda\rho_0^3}{|\alpha|^2}\label{26}
\end{equation}
Finally, the absorption cross section for the $l$-th partial wave
\cite{Gubser:1997qr} is given as a function of the frequency
$\omega$ and the quantized momentum $k$
\begin{equation}
\sigma^{(l\neq{3n})}_{abs}=2\pi^3(l+1)(l+2)(2l+3)A_{l\neq{3n}}^2r_{0}^{(2l+3)}\Big(\omega^2-k^2\Big)^{l-\frac{1}{2}}\Big(\frac{\omega}{T}\Big)\label{27}
\end{equation}
where
\begin{equation}
A_{l\neq{3n}}^2=\frac{|\Gamma(-\frac{2l}{3})\Gamma^2(1+\frac{l}{3})|^2}{|\Gamma(2+\frac{2l}{3})\Gamma^2(-\frac{l}{3})|^22^{(2l+3)}\Gamma^2(l+\frac{5}{2})}\label{28}
\end{equation}

The $l=3n$, $n=0,1,2,3,..., $ case is treated differently since the
regular at $x=1$ function $F_2(x)$ is expressed in terms of the
Legendre polynomials $P_n(x)$ as $F_2(x)\sim{P_n(\frac{2}{x^3}-1)}$.
Also, keep in mind that $F_2(x)\sim{x^{-l}}$ as $x$ goes to zero. In
this case the absorption cross section is
\begin{equation}
\sigma^{(l={3n})}_{abs}=2\pi^3(l+1)(l+2)(2l+3)\tilde{A}_{l=3n}^2r_{0}^{(2l+3)}\Big(\omega^2-k^2\Big)^{l-\frac{1}{2}}\Big(\frac{\omega}{T}\Big)\label{29}
\end{equation}
where
\begin{equation}
A_{l=3n}=\frac{1}{2^{(2l+3)}|\Gamma(l+\frac{5}{2})|^2}\label{30}
\end{equation}
Interestingly enough, one can calculate the s-wave cross section by
setting $l=0$ on Eq.~(\ref{28})
\begin{eqnarray}
\sigma_{abs}^{(l=0)}=2(\frac{4}{3})^{8}\pi^6R^9T^5\Big(\frac{\omega}{(\omega^2-k^2)^{\frac{1}{2}}}\Big)\sim{\frac{\tilde{A}_3}{V_3L}\Big(\frac{\omega}{(\omega^2-k^2)^{\frac{1}{2}}}\Big)}\label{31}
\end{eqnarray}
where $\tilde{A}_3$ stands for the area of the horizon.

Let us now comment on the expression of the absorption cross section
in the various cases we investigated. First of all, in all cases
$\sigma_{abs}$ depends not only on the frequency of the scalar wave
but also on the momentum along the smeared $U(1)$ direction. A
behavior like this was observed in the effective string models
\cite{Maldacena:1996ix,Gubser:1996zp} where Kaluza-Klein charged
fixed scalars are associated with the inclusion of a momentum along
a string. Also, similar results in black ring backgrounds have been
reported recently \cite{Cardoso:2005sj}. At this point we remind the
reader that the whole calculation was based on the assumption that
the energy of the scalar probe field and the momentum along the
smeared direction respect the following inequality $\omega^2>k^2$.
Given that, the absorption stays finite and at very low values of
momenta $k$ it acquires a fixed value which is independent of the
frequency in the s-wave approximation. However, when higher partial
waves are considered $\sigma\sim{\omega^{2l}}$ as easily seen in
Eq.~(\ref{27},\ref{29}).


A close inspection of Eq.~(\ref{31}) reveals that the cross section
probability is proportional to the area of the horizon times the
$\omega/(\omega^2-k^2)^{\frac{1}{2}}$ factor. We recall an important
result of Policastro, Son and Starinets \cite{Policastro:2001yc}
which states that for a non-extremal D$3$-brane the absorption cross
section in the zero frequency limit is exactly equal to the
horizon's area. This statement generalizes a previous known
universal result on black holes with spherically symmetric horizons
\cite{Das:1996we}. As is apparent, in our case though $\sigma_{abs}$
is not equal to the area of the horizon of the smeared D$3$-brane as
$\omega\rightarrow{0}$. This still holds even if $k=0$. We believe
that the intricate topology of the horizon (product of spheres) of
the smeared D$3$-brane plays a key role in the way the s-wave
absorption cross section shapes up.

\section{Cross Section And Temperature Corrections}
The absorption cross section calculation presented in the previous
section was related to a perturbative expansion Eq.~(\ref{17}) based
on which the zeroth order term $F^{(0)}$ was computed. We shall
attempt to compute the first order in $\lambda$ correction to the
cross section denoted by $F^{(1)}$ in the s-wave approximation. In
principle, one can go even beyond the first order by considering
higher corrections in $\lambda$ however, the complexity of the
resulted equations is daunting. In essence, the expansion we follow
is
\begin{equation}
F(x)=1+\lambda F^{(1)}(x)\label{32}
\end{equation}
Hence, it turns out that by substituting Eq.~(\ref{32}) directly
into Eq.~(\ref{16}) we get
\begin{equation}
\frac{d^2F^{(1)}}{dx^2}-\frac{2+x^3}{x(1-x^3)}\frac{dF^{(1)}}{dx}-\frac{i}{3}\Big(\frac{2+x^3}{x(1-x)(1-x^3)}-\frac{1}{(1-x)^2}\Big)=0\label{33}
\end{equation}
The solution to Eq.~(\ref{33}) which respects regularity at the
horizon reads
\begin{equation}
F^{(1)}(x)=-\frac{i}{3}\ln \Big(\frac{1+x+x^2}{3}\Big)\label{34}
\end{equation}
Finally, we conclude by providing the final expression of the
absorption cross section which reads
\begin{equation}
\sigma_{abs}^{(l=0)}=2(\frac{4}{3})^{8}\pi^6R^9T^5\Big(\frac{\omega}{(\omega^2-k^2)^{\frac{1}{2}}}\Big)\Bigg(1-\Big(\frac{\ln3}{4\pi}\Big)^2\Big(\frac{\omega}{T}\Big)^2+\cdots\Bigg)\label{35}
\end{equation}
 where the ellipsis stand for higher order temperature
corrections. In the above result we have tacitly assumed that
$(\omega \ln3/4\pi T)^2<<1$ which is in accordance with our
approximation scheme which vouchsafes for the positiveness of the
absorption cross section. To the next order in $\lambda$ the
differential equation which $F^{(2)}$ obeys reads
\begin{multline}
\frac{d^2F^{(2)}}{dx^2}-\frac{2+x^3}{x(1-x^3)}\frac{dF^{(2)}}{dx}
+\frac{2}{9}\Big(\frac{2x+1}{1-x^3}\Big)\\-\frac{1}{9}\frac{1}{(1-x)^2}+\frac{1}{x(1-x^3)^2}-\frac{1}{9}\ln\Big(\frac{1+x+x^2}{3}\Big)\Big(\frac{2+x^3}{x(1-x)(1-x^3)}-\frac{1}{(1-x^2)}\Big)=0\label{36}
\end{multline}
Even though we were not able to get the exact expression of
$F^{(2)}$ as a solution of Eq.~(\ref{36}), it turns out (through an
asymptotic analysis) that this function is regular at the horizon
and the boundary.

Although the effects of the smeared direction are nicely captured in
the expression of the absorption cross section it would be
interesting to study the effects of $k$ at higher order in the
perturbative expansion. For instance, it would be useful, if it is
possible, to compute $G^{(2)}(x)$ so that we get an idea of the
effects of the smeared direction on the problem at hand. It turns
out that $G^{(2)}$ is computable and regular at the horizon
\begin{equation}
G^{(2)}(x)=-\frac{r_0^2}{10}\ln
\Big(\frac{1+x+x^2}{3}\Big)-r_0^2\frac{\sqrt{3}}{15}
\tan^{-1}\Big(\frac{\sqrt{3}(2x+1)}{3}\Big)-\frac{r_0^2}{10}\Big(1-\frac{1}{x^2}-\frac{\sqrt{3}\pi}{45}\Big)\label{37}
\end{equation}
Obviously, the function is divergent as one approaches the boundary
($\rho\rightarrow{\infty}$). However, if one takes the double limit
$ x,r_0\rightarrow{0}$, $G^{(2)}$ becomes finite. Thus, the
irregularity at the boundary can be removed for small values of the
momentum and for ''tiny'' smeared black branes. We believe that the
fact that $G^{(2)}$ diverges at the boundary may signal the need to
impose different boundary conditions at infinity in place of the
usual Dirichlet conditions.


.


\section{Conclusions}
In this paper we fill a gap in the literature by presenting a
detailed calculation of the absorption cross section of a neutral
massless scalar field propagating in the vicinity of a black smeared
D$3$-brane. We basically started from the well known technique of
separating the spacetime (generated by the black brane) in several
asymptotic regions. This in particular enabled us to obtain an
explicit solution of the wave equation in each one of those regions.

One of the intricacies of our calculation was to develop a
perturbative scheme based on the two parameters $\lambda,k$ which
enter the calculation. Such an implementation is shown to work well
up to the first order in $\lambda$. The perturbative expansion on
the momentum $k$ was shown to be consistent with the boundary
conditions on the horizon. However, the expansion to the $k^2$ order
exhibits irregular behavior at infinity unless the black brane has
vanishingly small radius. Finally, as one approaches the
decompactification limit the wave function of the scalar field
should be expressed only in terms of a perturbative series in
$\lambda$. One might attribute those results to the non trivial
topology (product of spheres) of the horizon. To begin with, the
horizon has an $S^1\times{S^4}$ topology however as $
k\rightarrow{0}$ it will resemble more like to an $R^1\times{S^4}$.

Also, we pointed out that $\sigma_{abs}=\sigma(\omega, k)$. Hence,
$\sigma$ exhibits an additional dependance on the momentum along the
smeared direction, in contrast to what is known for ordinary black
d-branes. However we showed that for very small values of the
momentum $k$ along the smeared direction the absorption cross
section is directly proportional to the area of the horizon.
Finally, higher order temperature corrections to the absorption
cross section were carried out.

It would be interesting to pursue similar studies in other smeared
D-brane backgrounds so that we uncover useful information about
D-brane physics. It would be appealing to consider scalar
perturbations of massive scalars. This would certainly complicate
the calculations even further, however we might be able to see how
the absorption cross section is affected by the mass term. Finally,
one may also consider fermions probing the black brane and study
their effects in the low temperature regime as was done for scalars
in \cite{Policastro:2001yb}.

 \noindent
{\bf Acknowledgements}\\
I wish to thank Ioannis Bakas for his continuous guidance and for
useful discussions. I also thank Professor Niels Obers for an
illuminating correspondence.


\end{document}